\documentclass[twocolumn,
preprintnumbers,amsmath,amssymb,floatfix]{revtex4}
\usepackage{epsf}
\newcommand{\be}{\begin{equation}}
\newcommand{\ee}{\end{equation}}
\newcommand{\bea}{\begin{eqnarray}}
\newcommand{\eea}{\end{eqnarray}}
\newcommand{\bean}{\begin{eqnarray*}}

\newcommand{\eean}{\end{eqnarray*}}
\font\upright=cmu10 scaled\magstep1
\font\sans=cmss10
\newcommand{\ssf}{\sans}
\newcommand{\stroke}{\vrule height8pt width0.4pt depth-0.1pt}
\newcommand{\Z}{\hbox{\upright\rlap{\ssf Z}\kern 2.7pt {\ssf Z}}}

\newcommand{\C}{{\rlap{\rlap{C}\kern 3.8pt\stroke}\phantom{C}}}
\newcommand{\R}{\hbox{\upright\rlap{I}\kern 1.7pt R}}
\newcommand{\CP}{\C{\upright\rlap{I}\kern 1.5pt P}}
\newcommand{\PP}{\hbox{\upright\rlap{I}\kern 1.5pt P}}

\newcommand{\identity}{{\upright\rlap{1}\kern 2.0pt 1}}

\newcommand{\HH}{\mbox{\hbox{\upright\rlap{I}\kern 1.7pt H}}}

\newcommand{\fr}{\frac}

\newcommand{\ra}{\rightarrow}
\newcommand{\al}{\alpha}
\newcommand{\bt}{\beta}
\newcommand{\pr}{\partial}
\newcommand{\hs}{\hspace{5mm}}

\newcommand{\dg}{\dagger}

\newcommand{\zb}{{\bar z}}

\font\mybb=msbm10 at 11pt

\def\bb#1{\hbox{\mybb#1}}

\def\bC {\bb{C}}

\renewcommand{\CP}{\bC {\rm P}}

\begin{document}
\draft
\title{An Improved Harmonic Map Ansatz}
\author{Theodora Ioannidou}\email{ti3@auth.gr}
\affiliation{ Maths Division, School of Technology, University of
Thessaloniki, Thessaloniki 54124,  Greece}
\author{Burkhard Kleihaus}\email{kleihaus@theorie.physik.uni-oldenburg.de}
\affiliation{Institut f\"ur Physik, Universit\"at Oldenburg, Postfach 2503
D-26111 Oldenburg, Germany}
\author{
Wojtek Zakrzewski} 
\email{W.J.Zakrzewski@durham.ac.uk}
\affiliation{Department of Mathematical Sciences, University
of Durham, Durham DH1 3LE, UK}

\date{\today}

\begin{abstract}
The rational map ansatz of Houghton et al \cite{HMS}  is generalised
by allowing the profile function, usually a function of $r$, to depend
also on $z$ and $\bar{z}$.  It is  shown
that, within this ansatz, the energies of the lowest $B=2,3,4$ field
configurations of the $SU(2)$ Skyrme model are closer to the corresponding 
 values  of the true solutions of the model 
than those obtained within the original rational map ansatz.
In particular, we present plots of the profile functions which do exhibit 
their dependence on $ z$ and $\bar{z}$. 

The obvious generalisation of the ansatz to higher $SU(N)$ models involving the
introduction of  more projectors is briefly mentioned.\\
\end{abstract}

\maketitle

\renewcommand{\thefootnote}{\arabic{footnote}}
\setcounter{footnote}{0}
\renewcommand{\thefootnote}{\arabic{footnote}}
\setcounter{footnote}{0}

{\sl Introduction}\\

A few years ago Houghton et al \cite{HMS} presented an ansatz, the so-called
rational map ansatz, in order to approximate multiskyrmion solutions
of the $SU(2)$ Skyrme model 
 by field configurations in which the angular dependence 
was determined by a rational map and the radial dependence was determined 
by a numerical solution of a nonlinear ordinary differential equation. 
This last equation involved the so-called
profile function (i.e. the generalisation of the one skyrmion profile function)
 and its shape had to be determined numerically. 

This ansatz  has been a great step forward since it gave very good 
approximations to the solutions of the full equations which up to then 
could only
be determined numerically (and these simulations involved hours, days or weeks 
of CPU time). In fact, these approximations were so good that the values of 
the energies
were only, at most, a few \% up on the true value and it was practically
 impossible to distinguish the energy density plots obtained with the use 
of the ansatz  from the exact ones obtained numerically.

At the same time the ansatz clarified the situation for higher baryon 
number states; the energy density had maxima  at several points lying on a
 shell whose radius grew with the baryon number.
It has also lead to some generalisations - first, the generalisation to 
the $SU(N)$ Skyrme models lead to the harmonic map ansatz \cite{IPZ1}, 
in which the original rational map of Houghton et al \cite{HMS} was 
replaced by the projector of the more general harmonic map of the $CP\sp{N-1}$ 
model and then to the realisation that one could use more of
such projectors, which represent harmonic maps, to obtain further solutions 
(radial cases) or field configurations of the $SU(N)$ models. 
Since then, these ideas have been used to other models (i.e. monopoles) or 
similar ones coupled to gravity.

All these studies have relied on one fundamental assumption:
the separation of the angular degrees of freedom 
given in terms of a projector or a series of projectors 
and the radial dependence built in through the profile function 
or functions (for more projectors).

However, it is easy to check that the profile function does not have 
to depend only on $r$; it could also depend on the angular variables. 
As long as the projectors represent {\bf harmonic maps} - the profile
functions can be more general, and therefore provide  
a better approximation to the solutions of the model.
This observation constitutes the essence of our {\bf improved harmonic 
map ansatz}; the profile functions in addition to their dependence 
on $r$ do depend, also, on $z$ and $\bar{z}$ which allows for a more general
 angular dependence of the fields and density functions.

In this paper we present our ansatz and reinvestigate the lowest baryon 
number field configurations of the $SU(2)$ model. 
In particular we look at $B=2$, 3 and 4. 
In the  subsequent papers we will look at larger values of $B$; the 
$SU(N)$ models for $N>2$; and the  multiprojector case.
We will also look at the gravitating cases involving  skyrmions and monopoles.\\

{\sl The Harmonic Map Ansatz and its Improved Version}\\

 The $SU(N)$ Skyrme action to be considered is given by
 \be
  S=\int \left[\fr{\kappa^2}{4} \mbox{tr}\left(K_{\mu}\,K^\mu 
\right) +\fr{1}{32e^2}\mbox{tr}
\left(\left[K_\mu,K_\nu\right]\left[K^\mu,
K^\nu\right
]\right)\right]\, d^4x
  \label{ac}
  \ee
where $K_\mu=\pr_\mu U U^{-1}$ for $\mu=0,1,2,3$;  $U$ is the  $SU(N)$ 
chiral field and $\kappa$, $e$ are coupling constants. 
Although, we present the discussion for the general $SU(N)$ model 
in this paper we will give the details of our studies for the $SU(2)$ case.

 It is  well known that, in order for the finite-energy configurations to exist,
 the Skyrme field has to go to a constant matrix at spatial infinity:  
$U\ra I$ as $|x^\mu|\ra \infty$.
This effectively compactifies the three-dimensional Euclidean space
into $S^3$ and hence implies that the field configurations of the Skyrme
 model can be considered as maps from $S^3$ into $SU(N)$.

This compactification leads to the existence of a conserved topological current  
 yielding the topological charge to be identified with the baryon number 
$B$ defined as
\be
B=\int\,B^0\, d^3x
\label{b}
\ee
where 
\be
B^\mu=-\fr{1}{24\pi^2}\,\varepsilon^{\mu\nu\al\bt}\,\mbox{tr}\left
(K_\nu K_\al K_\bt\right)
\ee
and  $\varepsilon^{\mu\nu\al\bt}$ is the (constant) fully antisymmetric
tensor.

The starting point of the further discussion  is the introduction of the
coordinates $r,z,\zb$ on $\R^3$. In terms of the usual spherical
coordinates $r,\theta,\phi$ the Riemann sphere variable $z$ is defined by:
$z=e^{i\phi} \tan(\theta/2)$.
Then, the Skyrme action becomes \cite{HMS}
\bea
 S\!\!&=&\!\!\!\int drdt\,dz d\bar{z} \,\,
\mbox{tr}\bigg(\fr{\kappa^2r^2}{2(1+|z|^2)^2}
K_r^2+\frac{\kappa^2}{2}|K_z|^2\nonumber\\
&&+
\frac{1}{8e^2}\bigg|\left[K_r,K_z\right]\bigg|^2
-\frac{(1+|z|\sp2)^2}{32e^2r^2}
\left[K_z,K_{\bar{z}}\right]^2\bigg)
  \label{ac1}
  \eea
while the baryon number takes the form
\be
B=-\fr{1}{8\pi^2}\int\mbox{tr}\left(K_r\left[K_z,K_{\bar{z}}\right]\right)
 dr\, dz\,d\bar{z}.
\label{B}
\ee
Since we are interested in the static field configurations in what 
follows we  assume no $t$ dependence.

Next we consider the harmonic map ansatz which involves
assuming that the Skyrme field is  of the form
\cite{IPZ1}
\be
U=e^{2ih\left(P-I/N \right)}
=e^{-2ih/N}\left[I+\left(e^{2ih}-1\right)P\right]
\label{U}
\ee
where $P$ is a $N\times N$ hermitian projector which depends only on the
angular variables $(z,\bar{z})$ and $h$ is the  
profile function which depends, at least, on $r$.
Note that, the matrix $P$ can be thought of as describing a mapping 
from $S^2$ into $CP^{N-1}$ defined as 
\be
P(V)=\fr{V \otimes V^\dg}{|V|^2}
\label{for}
\ee
where $V$ is a $N$-component complex vector (depending on $z$ and
$\bar{z}$). For $N=2$ and $V=\left(1,f(z)\right)$ where $f(z)$ is a
 rational function
 we recover the rational map ansatz of 
Houghton et al \cite{HMS}.

Following \cite{DinZak}, we define a new operator $P_+$  by its action
on any vector $v \in \C^N$ as
\be
P_+ v=\pr_z v- \fr{v \,(v^\dg \,\pr_z v)}{|v|^2}.
\ee
Note that, when $V=V(z)$ (only) then
\be  P_z=\fr{P_+V \otimes V^\dg}{|V|^2}\ee
and so
\bea
 PP_z=0,\hs \hs&&  P_zP=P_z.\nonumber\\
 P_{\bar{z}}P=0,\hs \hs &&P P_{\bar{z}}=P_{\bar{z}},
\eea
where the subscripts denote partial derivatives.

For (\ref{U}) to be well-defined at the origin, 
the profile function   $h$, as a function of $r$, has to satisfy
$h(0)=\pi$ while the boundary value $U \ra I$ at $r=\infty$ requires that 
$h(\infty)=0$.
As shown in \cite{HMS, IPZ1},
an attractive feature of  (\ref{U}) is that it leads to a
simple expression for the energy density which can  be successively minimized
with respect to the parameters of the projector $P$ and then with respect
to the shape of the profile function $h$. 
This procedure for $h=h(r)$  gives good approximations to 
multiskyrmion field  configurations \cite{HMS,IPZ1}.

In what follows, 
this ansatz is ``improved" by allowing the profile
function $h$ to depend on $z$ and $\bar{z}$ in addition to
its $r$ dependence. 
The new ansatz is consistent with the partial factorisation 
of the field in the sense that
it still reduces the problem to having to solve
one equation for one function - namely $h$.
Had we taken an ansatz in which, say, the parameters of the projector $P$ 
depended on $r$, such a
simplification would not have taken place.
Thus, this modification is ``nontrivial"  and that is the reason for 
calling it an  {\it improved harmonic map ansatz}.
Note that,  $h$ has to be real 
 implying   that $h=h(r,|z|^2, {z+\bar{z}\over \vert z\vert})$; while, 
at the origin, we require that
 $h(0,|z|^2, {z+\bar{z}\over \vert z\vert})=\pi$.

The action (\ref{ac1}), due to  (\ref{U}) for  
$h=h(r,z,\bar{z})$
and  using  the aforementioned properties of the harmonic maps, becomes
\bea
S\!\!&=&\!\!\int\!\! dt\,dr\,dz\, d\bar{z}
\bigg(\!\!\!-\kappa^2\,A_N\, r^2\,h_r^2
-\kappa^2 B_N \,\vert h_z\vert^2\nonumber\\
&-&\!\!\left[{\cal N}_1\left(\kappa^2+\fr{h_r^2}{e^2}\right)
 +\fr{{\cal N}_2}{e^2}\fr{ \vert  h_z\vert^2}{r^2}\right]
\sin^2 h
-\fr{\cal I}{e^2}\fr{\sin^4 h}{r^2}\bigg)\nonumber\\
\label{s1}
\eea
 where
\bea
& &
A_N = 2i\,\fr{N-1}{N} \fr{1}{(1+|z|^2)^2} \ , \ \ \ 
B_N = 2i\,\fr{N-1}{N}\, \nonumber\\
& &
{\cal N}_1 = i\, 2\,\fr{|P_{+}V|^2}{|V|^2} \ , \ \ \ 
{\cal N}_2 = i 
\,\fr{|P_{+}V|^2}{|V|^2}\, (1+|z|^2)^2\,\nonumber\\
& &
{\cal I} = i\,\fr{|P_+V|^4}{|V|^4}\, (1+|z|^2)^2 
\eea
while the baryon number (\ref{B}) 
coincides with the expression for the topological  charge of the $CP^{N-1}$ 
sigma model (up to an overall profile dependent factor) since
\bea
B&=&\fr{i}{\pi^2}\int\mbox{tr}\left(P\left[P_z,P_{\bar{z}}\right]\right)
\,dz\,d\bar{z}
\int_{0}^{\infty} \sin^2 h \,h_r\,dr
\nonumber\\
&=&\fr{i}{2\pi}\int\,\fr{|P_+V|^2}{|V|^2}
\,dz\,d\bar{z}.
\label{b1}
\eea
It is easy to see that $\int_0\sp{\infty} \sin\sp2h\,h_r\,dr=h(r=0)/2
\equiv\pi/2$.

Next, for  convenience of our numerical simulations, 
we use spherical coordinates 
$(r, \theta , \phi)$ and introduce the dimensionless coordinate $x=e\kappa r$.
Variation of (\ref{s1}) with respect to  $h$  gives the equation of motion
\bea
&&\!\!\!\!\pr_x\left(\left[\frac{2(N-1)}{N}
+\frac{2\sin^2 h}{x^2} \,G \right]h_x\, x^2 \sin\theta \right)
\nonumber \\
&+&\!\!\pr_\theta\left(\left[\frac{2(N-1)}{N} +\frac{\sin^2 h}{x^2}\, G \right]
h_\theta\,\sin\theta \right)
\nonumber \\
&+&\!\!\pr_\phi\left(\left[\frac{2(N-1)}{N} + \frac{\sin^2 h}{x^2} 
\,G \right]
\frac{h_\phi}{\sin\theta} \right)
\nonumber \\
&-&\!\!\left(1+\fr{h_r^2}{2}+\frac{h_\theta^2}{2x^2}+\frac{h_\phi^2}{2x^2\,
\sin^2\theta}
           +\frac{\sin^2 h}{x^2} \,G
\right)\nonumber\\
&&\times  \sin(2 h) G \sin\theta =0	    
\label{ec}
\eea
where  $G=\fr{|P_+V|^2}{|V|^2}(1+|z|^2)^2$ is a function of $\theta$ and $\phi$ only.\\

{\sl $SU(2)$ $B=1, \dots , 4$ Baryons}\\

\begin{figure}
{\centerline{
\mbox{
\epsfysize=3.8cm
\epsffile{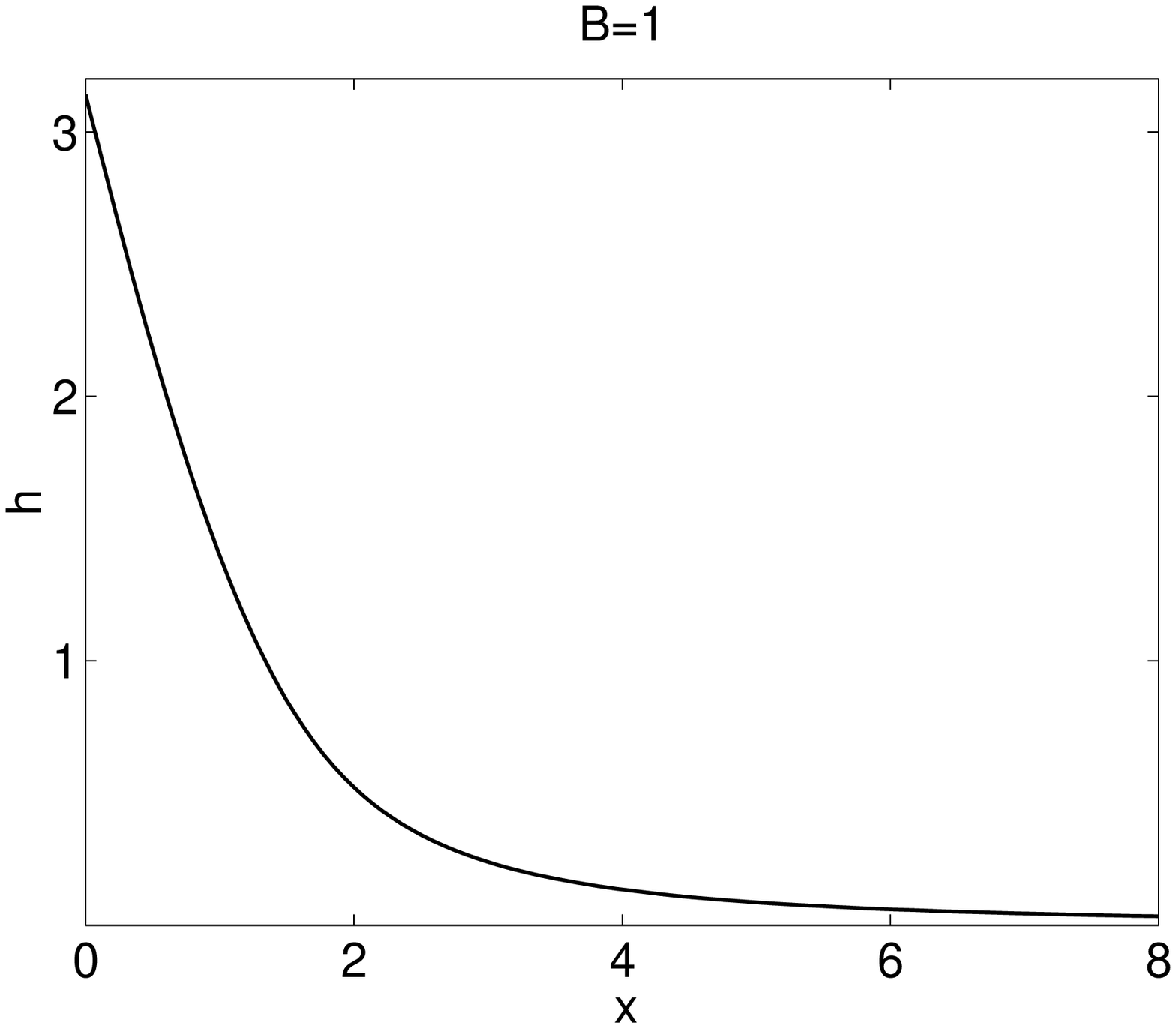}
\epsfysize=3.8cm
\epsffile{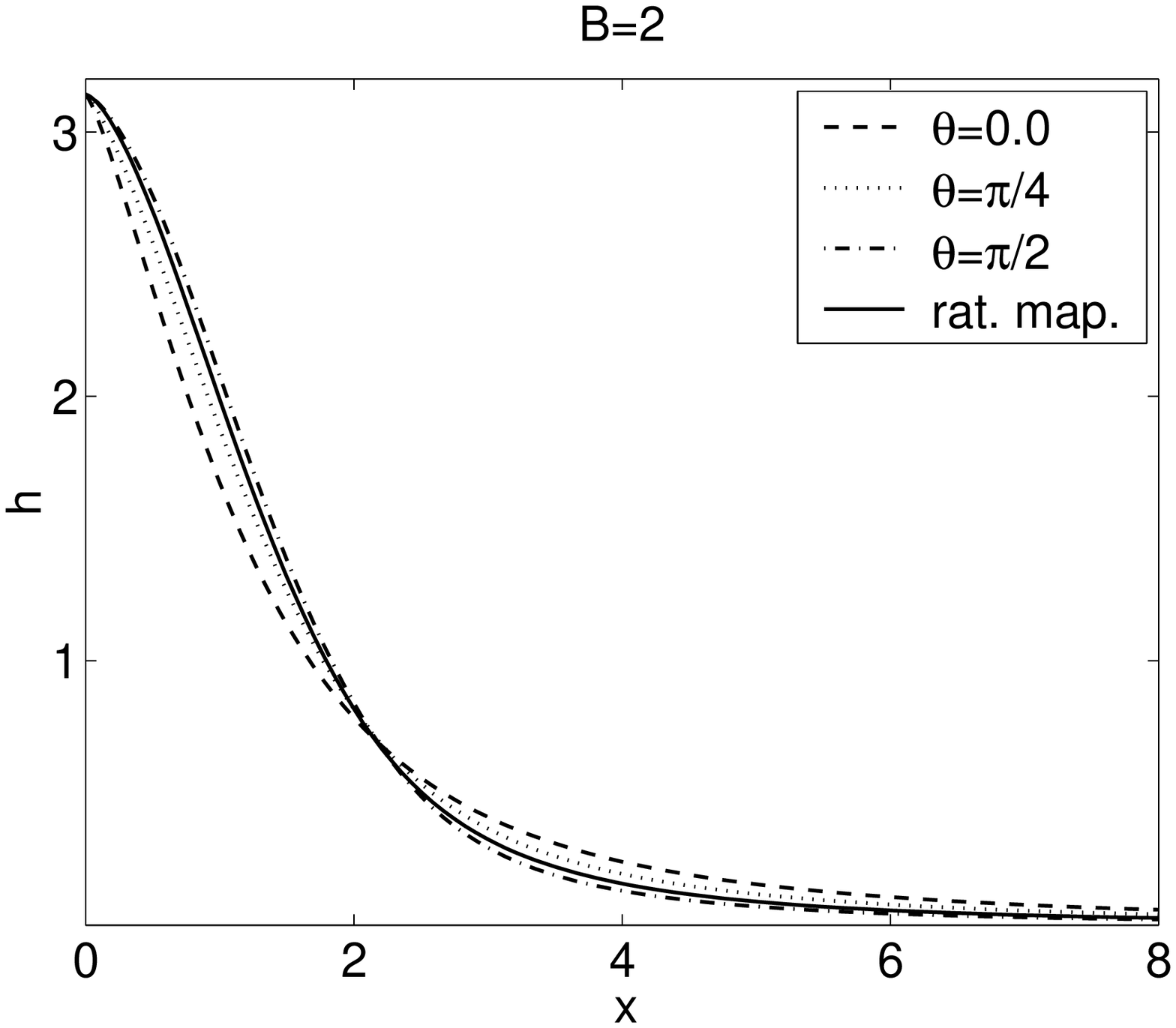}}}}
\vskip -.5cm
\caption{\label{f-1}The one skyrmion profile function showing
 no $\theta$ dependence (left) and
the two skyrmion profile function 
obtained from the improved harmonic and  rational map ansatz (right).
}
\end{figure}

First, to test our approach, we have calculated the profile
function $h$ for one skyrmion (i.e. for $B=1$) and found no
 $\theta$ or $\phi$ dependence  (as  expected), 
as Figure \ref{f-1} indicates. The total energy is 1.232.

\begin{figure}
{\centerline{
\mbox{
\epsfysize=3.8cm
\epsffile{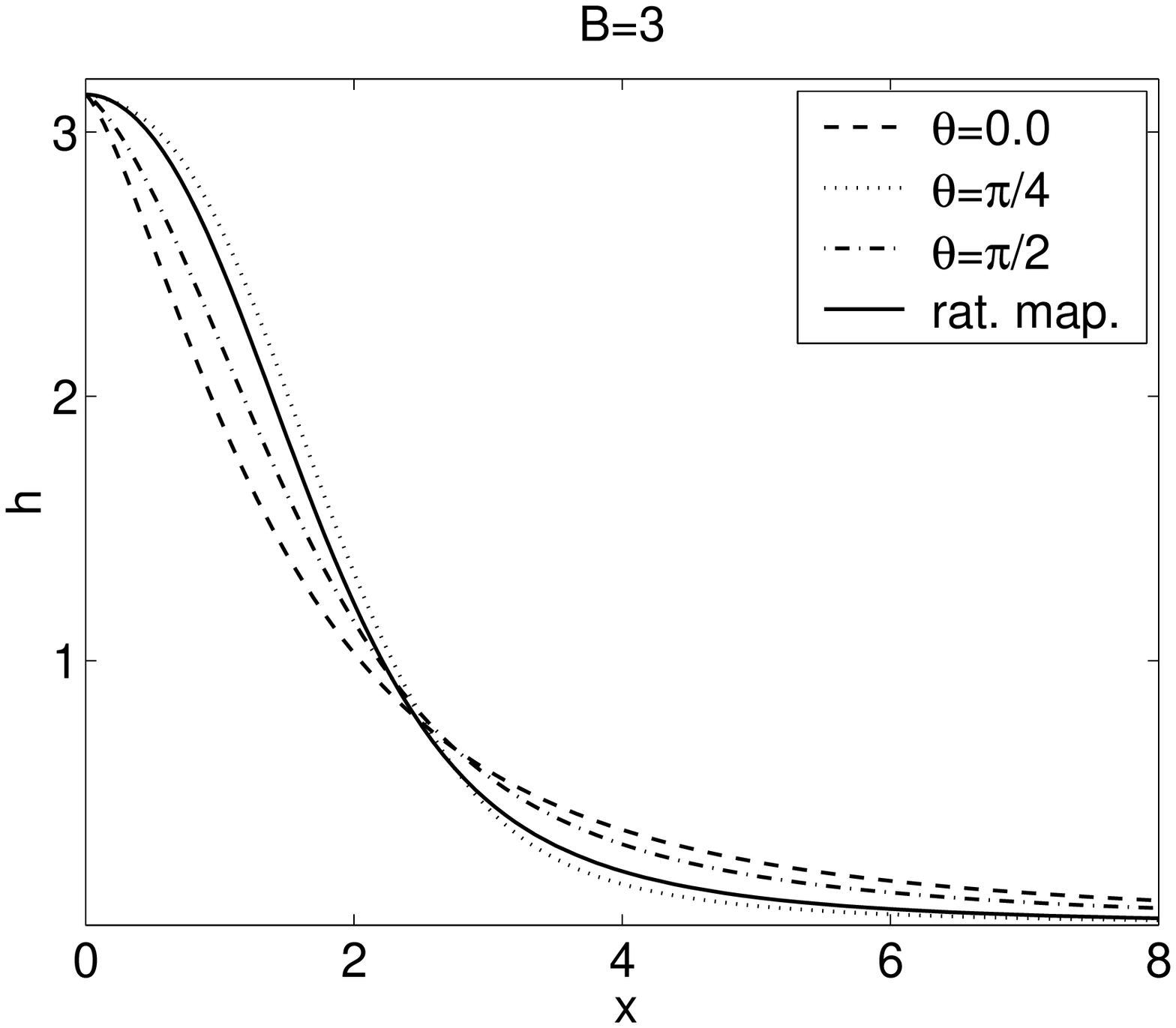}
\epsfysize=3.8cm
\epsffile{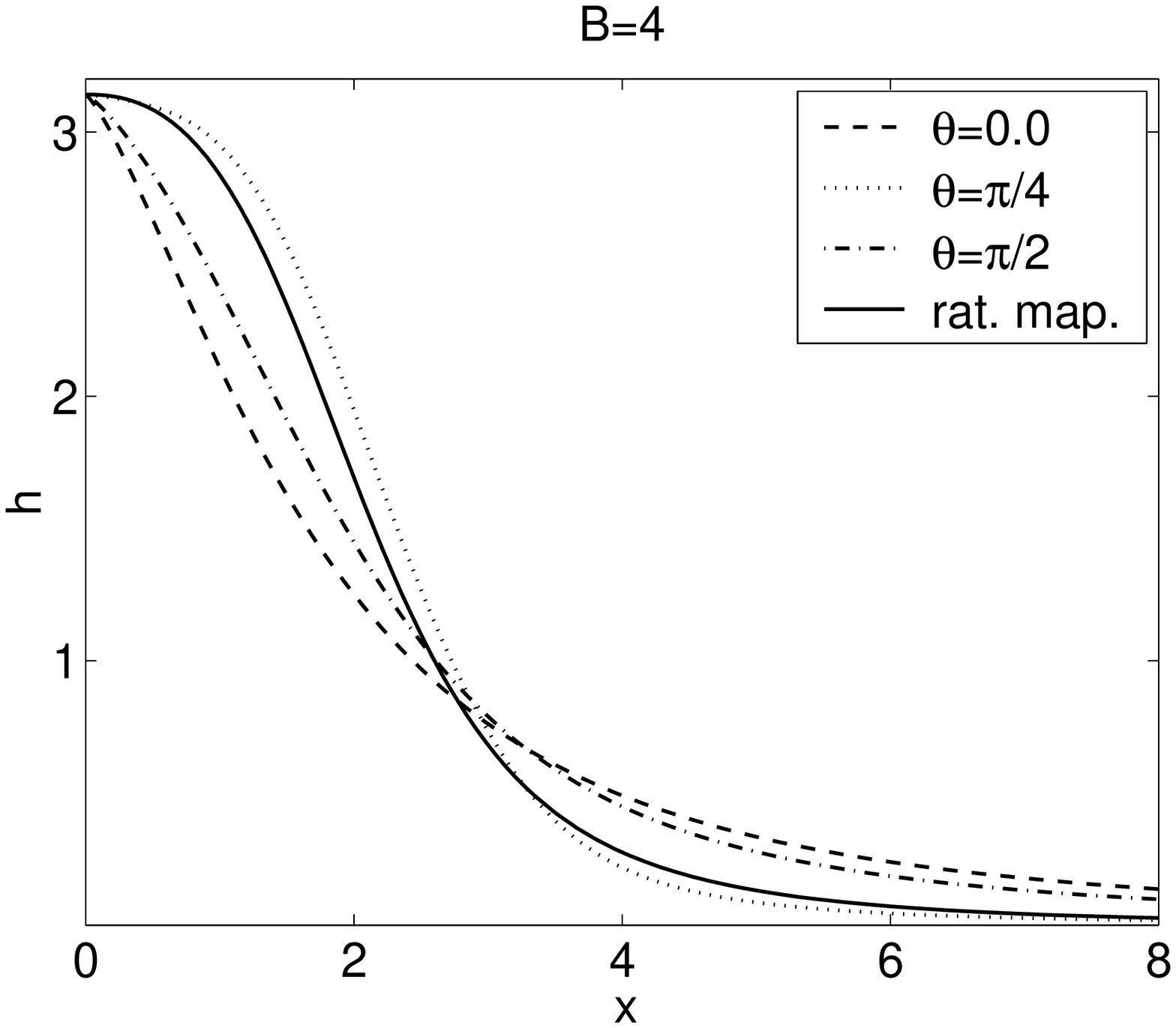}}}}\vskip -.5cm
\caption{\label{f-2} The skyrmion profile function $h(r,\theta)$
with axial symmetry
obtained from the improved harmonic  and  rational map ansatze 
for $B=3$ (left) and $B=4$ (right).}
\end{figure}

Axially symmetric skyrmions with baryon number $B>1$ can be 
obtained from vectors 
of the form \cite{HMS} 
\be
V=\left(z^B,1\right)^t.
\ee
In this case, the action (\ref{s1}) depends explicitly on $\theta$ 
but not on $\phi$. 
Consequently, we find $\theta$ dependence of the corresponding
profile functions $h$, as shown  in Figures \ref{f-1}  and \ref{f-2}.
The energies are given in Table 1 and compared with those 
obtained from the rational map ansatz.
For $B=2$ the energy is 2.382 which  is closer to 
the value 2.342 of the ``exact" solution (obtained numerically by solving
the full  equations \cite{BS}) and  lower than the value 2.416 obtained 
by the rational  map ansatz \cite{HMS}.
[In this case,  $h$ is $\phi$ independent (as expected)].
For $B>2$ the solutions (of this class)
 correspond to saddle points of the energy.
Obviously, the improved harmonic map ansatz yields lower energies than 
the rational map ansatz, as can be seen in Table 1.

Next, we consider skyrmions with platonic symmetry.
For $B=3$ and  $B=4$  we let the vector $V$ to be 
\bea
V&=&\left(\sqrt{3}iz\sp2-1,z(z\sp2-i\sqrt{3})\right)^t\nonumber\\
V&=&\left(z\sp4+2\sqrt{3}iz\sp2+1,z\sp4-2\sqrt{3}iz\sp2+1\right)^t.
\eea
All these expressions come from \cite{HMS} - who established their form by
minimizing the projector part of the action.
This time, the corresponding action (\ref{s1}) depends explicitly on 
$\theta$ and $\phi$. 
\begin{figure}
{\centerline{
\mbox{
\epsfysize=4.1cm
\epsffile{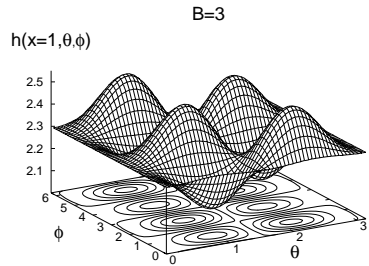}}}}
\vskip -0.5cm
\caption{\label{f-3} The $B=3$ skyrmion profile function
$h(r=1,\theta, \phi)$ with tetrahedral symmetry
obtained from the improved harmonic map ansatze.}
\end{figure}

First we consider the question whether we can obtain an improvement of 
the energy by considering  $h$ to depend (only) on $\theta$, in 
addition to $r$.
We have performed the integration of the action over the azimuthal 
angle $\phi$  which has given us an effective
action, which still contains an explicit $\theta$ dependence.
The variation of this effective action with respect to 
$h(r,\theta)$ led to  a partial differential equation for it.
We have found that the $\theta$ dependence of  $h$
is very small for $B=3$ while for $B=4$ it is more pronounced, but still small
compared to the axially symmetric solution.
As shown in Table 1, the improvement of the energy is rather small, 
compared to the improvement of the energy for the axisymmetric solutions.

\begin{figure}
{\centerline{
\mbox{
\epsfysize=4.1cm
\epsffile{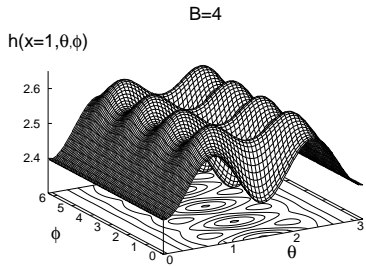}}}}
\vskip -0.5cm
\caption{\label{f-4} Same as Fig. \ref{f-3} for
$B=4$ with octahedral symmetry.}
\end{figure}
This result suggests that a considerable improvement
of the energy can only be achieved if we allow  $h$ to depend also 
on $\phi$, in addition to $r$ and $\theta$. 
The variational equation then yields a second 
order partial differential equation involving the independent 
variables $x, \theta, \phi$.
We solved this equation for $B=3$ and $B=4$ and found indeed 
a further improvement of the energies (see Table 1). 
For $B=3$, respectively,  $B=4$ the deviation from the  exact value is only 
$\approx$ 2 \%, respectively, $\approx$ 1 \%.
To demonstrade the angular dependence of the profile function we 
plot in Figures \ref{f-3}
and  \ref{f-4} $h(x,\theta,\phi)$ for fixed $x=e\kappa r=1$ for $B=3$ and $B=4$,
 respectively. 

\begin{widetext}
\begin{ruledtabular}
\begin{tabular}{|cccccccc|}
 \hline
\multicolumn{1}{|c}{}
& \multicolumn{3}{c}{axial symmetry} &\multicolumn{4}{c|}{platonic symmetry}
 \\
 \hline
$B$ &  rat. map & ``imp". harm. map & exact\cite{Paulpc}  & rat. map  & ``imp". harm. map (no $\phi$)& (with $\phi$) & exact\cite{BS}\\
 \hline
1 &  1.232    &   1.232       & 1.232  & --     & -      &  --      &  --   \\
2 &  1.208    &   1.191       & 1.181  & --     & -      &  --      &  --   \\
3 &  1.256    &   1.214       & 1.194  & 1.184  & 1.183  & 1.168    &  1.143\\
4 &  1.322    &   1.243       & 1.216  & 1.137  & 1.133  & 1.130    &  1.116
 \\
 \hline
\end{tabular}\vspace{7.mm}
\end{ruledtabular}
TABLE 1:
Energies per skyrmion  (i.e. $E/B$) of multiskyrmion configurations  obtained 
by the rational  and the  improved harmonic map ansatz (with and without 
$\phi$ dependence) for $B=1, \dots , 4$
 in comparison with the energies of the ``exact" solutions.\vspace{7.mm}
\end{widetext}

{\sl Conclusions}\\

In this letter we have pointed out that the rational map ansatz of 
Houghton et al \cite{HMS} can be further improved. 
The improvement involves the allowance of the profile function $h$
to depend on $z$ and $\bar{z}$ in addition to $r$.
 Of course, the rational map ansatz is already
a very good approximation; hence our improvement is only modest but, 
as we have shown (in the cases studied) the improvement is nonnegligible. 
To get our values we  first have assumed only $\theta$ dependence 
of $h$; however in order to get better values for the Skyrmions with 
platonic symmetries
we went further and allowed, also, its $\phi$ dependence.
Our modification of the ansatz
is not restricted to one projector. It is easy to see that if we want
to obtain low energy configurations of the $SU(N)$ model we could
use more projectors \cite{IPZ1} (up to $N-1$ for the $SU(N)$ case)
with the corresponding profile functions dependent on $r$, $\theta$
 and $\phi$.
We are also looking at such further applications of our ideas:
to systems involving $SU(N)$ Skyrme models for $N>2$ and to 
systems involving gravitational fields.

\end{document}